\documentclass[12pt,letterpaper]{article}
\usepackage[left=1in,top=1in,right=1in,nohead]{geometry}
\usepackage{enumerate}
\usepackage{graphicx}
\usepackage{amsmath}
\usepackage{amsthm}
\usepackage{mathrsfs}
\usepackage{amssymb}
\usepackage{multirow}
\usepackage{color}
\usepackage{bm}
\usepackage{comment}
\usepackage{subfigure}
\usepackage{paralist}
\usepackage{jheppub}

\newcommand{\T}{\mathcal{T}}
\newcommand{\Hbb}{\mathbb{H}}

\def\bea#1\eea{\begin{align}#1\end{align}}
\def\be#1\ee{\begin{equation}#1\end{equation}}
\def\dd{\mathrm{d}}

\title{Dissonant Black Droplets and Black Funnels}

\author[a]{Sebastian Fischetti,}
\author[b]{Jorge E. Santos,}
\author[c]{and Benson Way}

\affiliation[a]{Theoretical Physics Group, Blackett Laboratory, Imperial College, London SW7 2AZ, UK}
\affiliation[b]{DAMTP, Centre for Mathematical Sciences, University of Cambridge, Wilberforce Road, Cambridge CB3 0WA, UK}
\affiliation[c]{Department of Physics and Astronomy, University of British Columbia, 6224 Agricultural Road, Vancouver, B.C., V6T 1W9, Canada}

\emailAdd{s.fischetti@imperial.ac.uk}
\emailAdd{J.E.Santos@damtp.cam.ac.uk}
\emailAdd{benson@phas.ubc.ca}

\abstract{A holographic field theory on a fixed black hole background has a gravitational dual represented by a black funnel or a black droplet. These states are ``detuned'' when the temperature of the field theory near the horizon does not match the temperature of the background black hole.  In particular, the gravitational dual to the Boulware state must be a detuned solution. We construct detuned droplets and funnels dual to a Schwarzschild background and show that the Boulware phase is represented by a droplet. We also construct hairy black droplets associated to a low-temperature scalar condensation instability and show that they are thermodynamically preferred to their hairless counterparts.}

\begin{document}

\maketitle
\flushbottom

\section{Introduction}
\label{sec:intro}

The AdS/CFT correspondence~\cite{Mal97,Wit98a,GubKle98} provides a powerful means of studying strongly coupled quantum field theories in curved spacetimes.  Of particular interest are black hole backgrounds, as these enable the study Hawking radiation at strong coupling, a regime which would otherwise be intractable via standard field-theoretic means.  Moreover, since black holes act as heat sources and sinks, studying the behaviour of strongly-coupled field theories on black hole backgrounds can offer insight into heat transport in such theories.

\begin{figure}[t]
\centering
\subfigure[]{
\includegraphics[width=0.3\textwidth,page=3]{Figures-pics}
\label{subfig:funnel}
}
\hspace{0.1cm}
\subfigure[]{
\includegraphics[width=0.3\textwidth,page=2]{Figures-pics}
\label{subfig:generaldroplet}
}
\hspace{0.1cm}
\subfigure[]{
\includegraphics[width=0.3\textwidth,page=1]{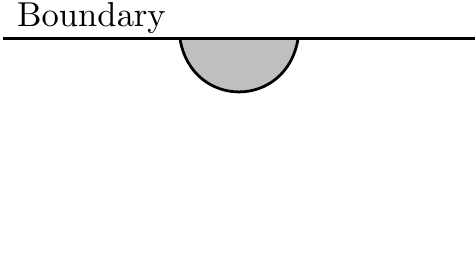}
\label{subfig:Unruhdroplet}
}
\caption{Sketches of black droplets and black funnels for an asymptotically flat boundary black hole.  From left to right, a black funnel, a black droplet with ~$T_\infty \neq 0$, and a black droplet with~$T_\infty = 0$.  The~$T_\infty = 0$ droplet can be thought of as the limit~$T_\infty \to 0$ in which the planar horizon becomes a Poincar\'e horizon.}
\label{fig:dropletfunnels}
\end{figure}

Let us therefore consider holographic CFTs with a background metric that contains a black hole, a programme first initiated in~\cite{HubMar09} and continued in~\cite{HubMar09b,CalDias11,FigLuc11,FisMar12,SanWay12,FisMar12b,FigWis12,FisSan13,FigTun13,Haddad13,EmpMar13,SanWay14,EmpShi15}. (See \cite{MarRang14} for a review.)  In the limit in which the bulk is described by classical general relativity, the gravitational duals are asymptotically locally AdS spacetimes (with AdS scale~$\ell$) which can roughly be divided into two classes: ``black funnels'' and ``black droplets''.  The distinction between these two classes is made based on the connectedness of the bulk horizon(s), as shown schematically in Figure~\ref{fig:dropletfunnels} for an asymptotically flat boundary black hole.

Generically, a fixed boundary geometry admits a family of bulk duals consisting of both black funnels and black droplets, with each bulk solution corresponding to a different state of the boundary field theory.  These states can be characterised by two dimensionless parameters\footnote{We assume that a notion of local thermodynamic equilibrium exists at the black hole horizon and at infinity.  We also assume that the boundary black hole is nonextremal so that~$T_\mathrm{BH}\neq 0$.}:
\begin{itemize} 
	\item The ratio~$\T_H \equiv T_H/T_\mathrm{BH}$ of the temperature~$T_H$ of the field theory near the black hole horizon to the black hole temperature~$T_\mathrm{BH}$; and
	\item The ratio~$\T_\infty \equiv T_\infty/T_\mathrm{BH}$ of the temperature~$T_\infty$ of the field theory in an asymptotic region (assuming one exists) to the black hole temperature.
\end{itemize}
The temperatures~$T_H$ and~$T_\infty$ determine the temperature of the bulk horizon(s) where they intersect the boundary and at an asymptotic region, respectively.  In contrast, since the boundary metric is nondynamical, the boundary black hole temperature~$T_\mathrm{BH}$ should instead be thought of as a geometrical scale.  In particular, there is no need for $T_\mathrm{BH}$ to match the bona fide thermodynamic field theory temperature~$T_H$. That is, it is permissible to take~$\T_H \neq 1$, in which case we call the resulting solution ``detuned''.

From the perspective of the CFT dual to these detuned solutions, the Euclidean path integral has a period~$\beta=1/T_{H}$, which means the Euclidean boundary geometry exhibits a conical singularity at the (boundary black hole) horizon.  This singularity is a consequence of the fact that the CFT is not in equilibrium with the black hole, and can be thought of as arising from an infinitesimally thin heat bath at temperature~$T_{H}$ just outside the horizon.  As a result, the stress energy tensor will diverge there.

Our purpose in this paper is to explore the space of detuned states.  We are particularly interested in the Boulware state, defined here as the lowest-energy state on a given boundary black hole\footnote{Other interesting states that are not detuned include the Hartle-Hawking state, characterized by~$\T_H = \T_\infty = 1$ and constructed in~\cite{SanWay12}, and the Unruh state, characterized by~$\T_H = 1$ and~$\T_\infty = 0$ and explored in~\cite{FigLuc11}.}.  We will work in four boundary dimensions, placing the field theory on a Schwarzschild black hole.  It is reasonable to expect that the Boulware state should have~$\T_H = \T_\infty = 0$, and indeed, we will construct a droplet with~$\T_H = \T_\infty = 0$ and show that it has a lower energy than any other known solution on the same background\footnote{We point out that droplets with extremal boundary black holes have been constructed~\cite{FisSan13,FigTun13,Mef16}.  The extremal case satisfies $T_{\mathrm{BH}}=T_\infty=T_{H}=0$, which can be thought of as simultaneously all of the Hartle-Hawking, Unruh, and Boulware vacua.}.  Moreover, we also attempt to construct a funnel with~$\T_H = \T_\infty = 0$, but our results suggest that a finite-temperature funnel ``pinches off'' to a droplet before zero temperature can be reached, providing strong evidence that the bulk dual to the Boulware state is a droplet.

While most of the geometries presented in this paper are solutions to the vacuum Einstein equation (with negative cosmological constant), we note that droplet and funnel geometries exist even in the presence of matter.  Moreover, at low temperatures, it is possible for such solutions to dominate the thermodynamic ensemble over the pure vacuum solutions.  For instance, a scalar field of sufficiently low mass will condense to form scalar hair around a black hole with small enough temperature (the most well-known example of this phenomenon may be the holographic superconductor~\cite{HarHer08}).  Because here we are primarily interested low temperatures, such solutions are relevant in our characterization. 

To briefly review the mechanism by which this condensation occurs, consider introducing a scalar field of mass~$\mu$ in the D-dimensional bulk.  At low temperature, the near-horizon geometry of the bulk black hole approaches~$\mathrm{AdS}_2 \times X^{D-2}$ for some transverse space~$X$.  Thus if the mass of the scalar is chosen such that~$\mu^2 \ell^2 < -1/4$, so that it violates the Breitenl\"ohner-Freedman (BF)~\cite{BreFre82} bound of the near-horizon~AdS$_2$\footnote{Note that the mass must still obey the BF bound~$\mu^2 \ell^2 \geq -(D-1)^2/4$ imposed by the AdS$_D$ asymptotics.}, an instability develops causing the scalar field to condense around the horizon.  This hairy solution is thermodynamically preferred, and thus a theory containing such a bulk matter field will have a different Boulware state from one without it.  To illustrate this phenomenon explicitly, we will therefore introduce a scalar field~$\varphi$ and construct black droplets which are solutions to the Einstein-Klein-Gordon system
\be
\label{eq:einstein}
E_{ab} \equiv R_{ab} +\frac{4}{\ell^2}g_{ab}-2\left(\nabla_a\varphi \nabla_b\varphi+\frac{1}{3}\mu^2\varphi^2g_{ab}\right)=0\;,\qquad(\nabla^2-\mu^2)\varphi=0,
\ee
where we take the mass of the scalar field to be~$\mu^2=-4/\ell^2$, which saturates the~AdS$_5$ BF bound.  As expected, we will find that at sufficiently low temperature, the hairy black droplets have lower energy than their hairless counterparts.

This paper is organized as follows.  In Section~\ref{sec:construction}, we outline the construction of our Boulware droplets (both with and without scalar hair), reviewing boundary conditions and numerical methods.  In Section~\ref{sec:constructionbf}, we outline the construction of (hairless) black funnels with~$\T_H = \T_\infty \neq 1$, and show evidence that as these temperatures are decreased, the funnels become singular and ``pinch off'' before zero temperature is reached.  In Section~\ref{sec:stresstensor} we examine the stress tensors of our droplet and funnel solutions, showing that the zero-temperature droplet has minimum energy.  We conclude in Section~\ref{sec:discussion} with a discussion on negative energies and future directions.


\section{Constructing Droplets}
\label{sec:construction}

Let us outline the construction of a Boulware droplet.  The methods used in this construction are commonly used in the literature, so we will relegate details to the Appendices.

\subsection{The Geometry of Black Droplets}
\label{subsec:geometry}

The geometry of a droplet with~$T_\infty = 0$ is shown schematically in Figure~\ref{subfig:Unruhdroplet}.  The boundary metric of this solution is conformal to Schwarzschild, 
\be
\label{eq:schwarzschild}
\dd s^2_{\mathrm{Schw}}= -f(r)\dd t^2+\frac{\dd r^2}{f(r)}+r^2 \dd\Omega_2^2, \mbox{ where } f(r) \equiv 1-\frac{r_s}{r}
\ee
and~$r_s$ is the Schwarzschild radius, which sets the boundary black hole temperature as $T_{\mathrm{BH}}=1/(4\pi r_s)$.  The $SO(3)$ symmetry of the sphere extends into the bulk, which contains an axis corresponding to a fixed point of this symmetry.  The bulk droplet horizon has temperature $T_{H}$ and extends from this axis to the boundary.  Finally, the solution ends on a Poincar\'{e} horizon, which we will write in the following way.  Beginning with~$\mathrm{AdS}_5$ in Poincar\'{e} coordinates,
\be
\dd s^2_\mathrm{AdS_5}=\frac{\ell^2}{z^2}\left(-\dd t^2+\dd z^2+\dd R^2+R^2 \dd\Omega_2^2\right),
\ee
we convert to algebraic polar coordinates $\rho = 1/\sqrt{R^2+z^2}$ and $\chi = R/\sqrt{R^2+z^2}$, in which case the metric becomes
\be
\label{eq:poincare}
\dd s^2_\mathrm{AdS_5} = \frac{\ell^2}{1-\chi^2}\left(-\rho^2\dd t^2+\frac{\dd\rho^2}{\rho^2}+\frac{\dd\chi^2}{1-\chi^2}+\chi^2\dd\Omega_2^2\right).
\ee
In these coordinates, the Poincar\'e horizon lies at~$\rho=0$ and its extremal nature is explicit.

\begin{figure}[t]
\centering
\subfigure[]{
\includegraphics[width=0.25\textwidth,page=4]{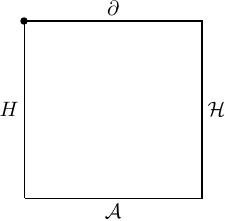}
\label{subfig:Unruhdomain}
}
\hspace{3cm}
\subfigure[]{
\includegraphics[width=0.25\textwidth,page=5]{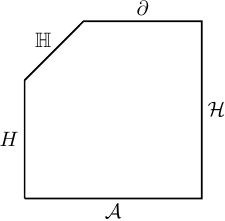}
\label{subfig:Boulwaredomain}
}
\caption{\subref{subfig:Unruhdomain}: The domain of integration for constructing the Unruh droplet.  The four boundaries correspond to the conformal boundary~$\partial$, the bulk horizon~$H$, the axis of symmetry~$\mathcal{A}$, and the Poincar\'e horizon~$\mathcal{H}$.  Because the bulk and boundary horizons have the same temperature ($\T_H = 1$), the coordinate ``point'' where the bulk horizon meets the boundary is well-behaved.  \subref{subfig:Boulwaredomain}: For a non-Unruh droplet ($\T_H \neq 1$), the metric is ill-behaved at this point.  To resolve this singular behavior, the point can be blown up into an asymptotically hyperbolic region~$\Hbb$, yielding a five-sided integration domain.}
\label{fig:dropletdomains}
\end{figure}

The special case of the Unruh droplet (in which~$\T_H = 1$) was constructed numerically in~\cite{FigLuc11}.  In such a case, the bulk horizon has the same temperature as the boundary black hole, and the coordinate ``point'' where they join is well-behaved.  This solution was found numerically by the Einstein equation on a rectangular domain whose four boundaries correspond to the conformal boundary, the bulk horizon, the axis of symmetry, and the Poincar\'e horizon; see Figure~\ref{subfig:Unruhdomain}.

However, if~$\T_H \neq 1$, this point will become a multi-valued coordinate singularity (the value of metric components there will depend on the direction from which it is approached), which must be more carefully resolved.  Essentially, this is accomplished by following the approach described in detail in~\cite{FisMar12} and used again in~\cite{SanWay12}, wherein a coordinate transformation is used to expand this ``point'' into an additional asymptotic region which approaches a hyperbolic AdS-Schwarzschild black hole of temperature~$T_H \neq T_\mathrm{BH}$; see Figure~\ref{subfig:Boulwaredomain} for an illustration and Appendix~\ref{app:detuning} for more details.  After transforming this singular point to an asymptotic region, the domain now consists of \textit{five} boundaries: the conformal boundary, a hyperbolic black hole, the bulk horizon, the axis of symmetry, and the Poincar\'e horizon.


\subsection{Ansatz}
\label{subsec:ansatz}

\begin{figure}[t]
\centering
\includegraphics[width=0.35\textwidth,page=17]{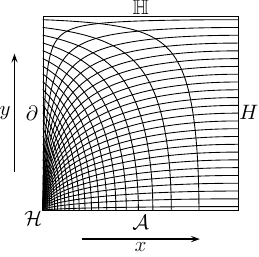}%
\hspace{1cm}
\includegraphics[width=0.35\textwidth,page=18]{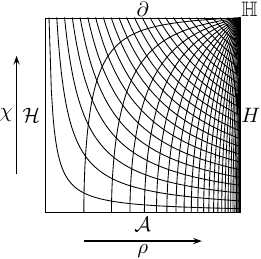}
\caption{The two coordinate systems used to construct detuned droplets.  The left figure shows the~$(x,y)$ coordinate square, within which we show the (distorted)~$(\rho,\chi)$ coordinate grid.  Likewise, the right figure shows the~$(\rho,\chi)$ coordinate square with the~$(x,y)$ coordinate grid drawn inside.  We have labeled the five boundaries of the computational domain; it is clear that the~$(x,y)$ coordinates are regular everywhere except at the Poincar\'e horizon~$\mathcal{H}$, while the~$(\rho,\chi)$ coordinates are regular everywhere except at the hyperbolic black hole~$\mathbb{H}$.}
\label{fig:patching}
\end{figure}

Let us now present our ansatz for detuned droplets.  Because the computational domain is five-sided, it is convenient to introduce two separate coordinate systems, each of which is regular on four of the five boundaries.  We will call one coordinate system~$(x,y)$, which is regular everywhere except at the Poincar\'e horizon, while we will call another~$(\rho,\chi)$, regular everywhere except on the hyperbolic black hole.  All of these coordinates range within in the unit interval, and they are related by the transformations
\be
\label{eq:coordtrans}
x = 1-\sqrt{-\frac{1-\alpha}{2\alpha}+\sqrt{\left(\frac{1-\alpha}{2\alpha}\right)^2+\frac{(1-\rho^2)^2}{\alpha(1-\chi^2\rho^2)^2}}}, \qquad y = \chi\rho,
\ee
where~$\alpha$ is a parameter that determines the detuning as
\be
\label{eq:hypertemp}
\T_H = \frac{1-\alpha}{\sqrt{1+\alpha}}.
\ee
See Figure~\ref{fig:patching} for an illustration of these coordinate systems.

In the~$(x,y)$ coordinates, our ansatz then takes the form\footnote{We do not have a straightforward way of motivating this particular ansatz. Its design is partially based on taking equation (2.14) in~\cite{FigLuc11} which is a reference metric for the Unruh droplet that can be understood as written in polar coordinates, and performing a map to a bipolar coordinate system.}
\begin{multline}
\label{eq:ansatz}
\dd s^2 =\frac{\ell^2}{x(2-x)(1+\alpha)}\left[-\frac{64(1-x)^2hF^2G^4Q_1\,\dd t^2}{r_s^2H^8}+\frac{(1+\alpha)Q_2\,\dd x^2}{x(2-x)h} \right. \\
\left. + \frac{4Q_3}{(2-y^2)(1-y^2)^2}\left(\dd y-\frac{y(1-y^2)(1-x)Q_5\,\dd x}{F}\right)^2+\frac{y^2(2-y^2)Q_4 \dd\Omega_2^2}{(1-y^2)^2}\right],
\end{multline}
where
\begin{subequations}
\bea
h &= 1-\alpha x(2-x), \\
F &= 1-h(1-x)^2(1-y^2)^2, \\
G &= 1+\sqrt{h}(1-x)(1-y^2), \\
H &= G+\sqrt F,
\eea
\end{subequations}
and the $Q_i$ are unknown functions of $x$ and $y$.

The boundary conditions on the~$Q_i$ are described in detail in Appendix~\ref{subapp:droplet}.  The case with the scalar field turned off ($\varphi = 0$) is simplest:
\begin{itemize}
	\item At the conformal boundary ($x = 0$,~$\chi = 1$), hyperbolic region ($y = 1$,~$\rho = \chi = 1$), and Poincar\'e horizon ($\rho = 0$,~$x = y = 0$), we impose the Dirichlet conditions $Q_1=Q_2=Q_3=Q_4=1$ and $Q_5=0$.
	\item At the bulk horizon ($x = 1$,~$\rho = 1$) we impose regularity, requiring~$\partial_x Q_i = 0$ and~$Q_1 = Q_2$.
	\item At the axis of symmetry ($y = 0$,~$\chi = 0$), we impose regularity, requiring~$\partial_y Q_i = 0$ and~$Q_3 = Q_4$.  In the~$(\rho,\chi)$ coordinates, these become~$Q_3=Q_4$ and $\partial_\chi Q_i=0$.
\end{itemize}

With the scalar field turned on, the black hole acquires scalar hair, which in particular means that the hyperbolic black hole does as well.  Thus the boundary condition at the asymptotically hyperbolic region becomes modified.  To obtain it, we first obtain the hyperbolic hairy black hole in a gauge\footnote{Specifically, in the De Turck gauge, described in the following section.} that is compatible with our ansatz \eqref{eq:ansatz}.  For that purpose, we look for hairy hyperbolic black holes of the form
\begin{multline}
\label{eq:hairyhyperbolic}
\dd s^2_\Hbb = \frac{\ell^2}{x(2-x)(1+\alpha)}\left[-\frac{(1-x)^2(1-\alpha x(2-x))Q_{\Hbb 1}}{4r_s^2} \, \dd t^2 \right. \\ \left. + \frac{(1+\alpha)Q_{\Hbb 2}}{x(2-x)(1-\alpha x(2-x))} \, \dd x^2 \right. \\ \left. + Q_{\Hbb 3}\left(\frac{4y^2\,\dd y^2}{(1-y^2)^2(1+(1-y^2)^2)}+\frac{\dd\Omega_2^2}{(1-y^2)^2}\right)\right],
\end{multline}
where now the $Q_{\Hbb i}$ and the scalar field~$\varphi_\Hbb$ are functions of $x$ only which must be obtained numerically.  For boundary conditions, we have $Q_{\Hbb i}=1$, $\varphi_\Hbb=0$ on the boundary~$(x=0)$, and $\partial_x Q_{\Hbb i}=0$, $\partial_x\varphi_\Hbb=0$ on the horizon~$(x=1)$.  The resulting Einstein equation (when written in De Turck form; see the following subsection) yields a set of ordinary differential equations (ODEs) for the $Q_{\Hbb i}$ which we solve numerically.  The solution of these ODEs reproduces the solution in \cite{DiaMon10}, but in a different gauge.  The critical temperature at which hairy hyperbolic black holes form corresponds to $\T_{\mathrm H}\approx 0.0339$.

To obtain the hairy droplet solution, we then continue to use the ansatz~\eqref{eq:ansatz} with the same boundary conditions except for the hyperbolic region ($y = 1$, $\rho = \chi = 1$), where we require that~\eqref{eq:ansatz} approach~\eqref{eq:hairyhyperbolic}: $Q_1=Q_{\Hbb 1}$, $Q_2=Q_{\Hbb 2}$, $Q_3=Q_4=Q_{\Hbb 3}$, and $Q_5=0$.  Moreover, on the scalar field we impose~$\varphi=\varphi_\Hbb$ at the hyperbolic region,~$\varphi=0$ at the conformal boundary and Poincar\'e horizon,~$\partial_y\varphi=0$ on the axis, and~$\partial_x \varphi=0$ on the horizon.


\subsection{Numerical approach}
\label{subsec:numerics}

Our numerical approach is the Einstein-de Turck method, which was first introduced in \cite{HeadKitch10,Wise11}, and reviewed in some detail in e.g.~\cite{DiaSan15}.  This approach requires a choice of reference metric~$\overline{g}$ which obeys the same boundary conditions as the desired solution~$g$, but which otherwise may be freely specified.  Once such a reference metric is supplied, one solves the Einstein-de Turck equation, which modifies the Einstein equation in~\eqref{eq:einstein} to 
\be
\label{eq:einsteindeturck}
E_{ab}-\nabla_{(a}\xi_{b)}=0, \mbox{ with } \xi^a \equiv g^{bc}(\Gamma^{a}_{bc}-\overline\Gamma^{a}_{bc}),
\ee
where~$\Gamma$ and~$\overline \Gamma$ define the Levi-Civita connections for the metric $g$ and reference metric $\overline{g}$, respectively.  In order for a solution to the Einstein-de Turck equation to also be a solution to unmodified Einstein equation, one requires the so-called de Turck gauge condition~$\xi^a=0$. Solutions with $\xi^a\neq0$ have been shown not to exists if one demands stationarity and the so-called $t-\phi$ reflection symmetry \cite{FigLuc11,FigWise16}, which are obeyed by both black funnels and droplets.  We will therefore use~$\xi^2$ (which must vanish in the continuum limit) to monitor the convergence of our solutions.

As a reference metric for both hairy and non-hairy droplets, we choose~\eqref{eq:ansatz} with $Q_1=Q_2=Q_3=Q_4=1$ and $Q_5=0$.  Since the equations of motion do not depend on $\ell$ or $r_s$, this is a one-parameter family of metrics parametrised by $\alpha$.  Similarly, as a reference metric for the hairy hyperbolic black holes, we choose~\eqref{eq:hairyhyperbolic} with~$Q_{\Hbb i}=1$.  

Next, we must properly control the five boundaries in the computational domain.  To that end, we employ patching, which was previously used to find black droplet solutions with bulk planar black holes~\cite{SanWay14}.  The idea is to divide the integration domain into two non-overlapping patches, one in the~$(x,y)$ coordinates and one in the $(\rho,\chi)$ coordinates (c.f.~Figure~\ref{fig:patching}).  We choose the patch boundary to extend from where where the horizon meets the axis to somewhere in the middle of the conformal boundary.  Unlike the approach in~\cite{SanWay14}, we do not define new metric functions in the $(\rho,\chi)$ coordinate system because this patch is mainly used to control the Poincar\'{e} horizon, which is a simple Dirichlet boundary condition.  We additionally impose continuity of the metric functions and their first derivatives across the patch boundary.

We then solve the elliptic equations~\eqref{eq:einsteindeturck} (subject to the aforementioned boundary conditions) using a standard Newton-Raphson algorithm.  We find that the reference metric supplies a suitable seed for any of these solutions.  The equations on the two patches are discretised by pseudo-spectral collocation methods using a tensor product of Chebyshev-Gauss-Lobotto nodes, which conform to the patches via transfinite interpolation.  The resulting algebraic system is solved using LU decomposition.

For any analytic function, pseudo-spectral collocation predicts exponential convergence with increasing grid size.  Indeed, we see this behaviour for our nonzero temperature ($\alpha \neq 1$) solutions in Figure~\ref{fig:convergence}.  However, as described in Appendix~\ref{app:conv}, in the extremal ($\alpha = 1$) case our solutions develop non-analytic behaviour at the hyperbolic black hole~$\mathbb{H}$.  This non-analyticity spoils the exponential convergence of the pseudo-spectral collocation, but it is nevertheless possible to determine the expected convergence given the particular non-analyticity.  As we show in Appendix~\ref{app:conv}, for an~$(N+N) \times N$ grid, this expected convergence goes like~$N^{-(\sqrt{11}-1)} \approx N^{-2.3}$, which is close to the~$\sim N^{-2.5}$ behaviour we observe.

\begin{figure}
\centering
\subfigure[]{
\includegraphics[width = 0.45\textwidth,page=9]{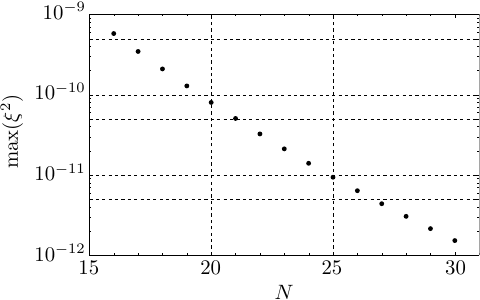}
\label{subfig:convergenceexp}}
\subfigure[]{
\includegraphics[width = 0.49\textwidth,page=10]{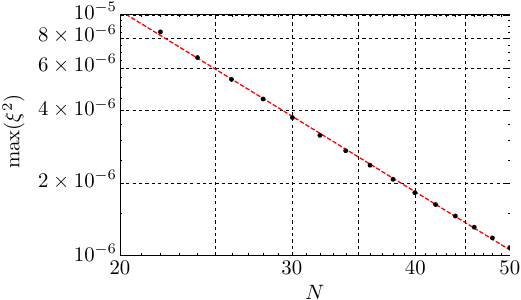}
\label{subfig:convergencepower}}
\caption{\subref{subfig:convergenceexp}: Convergence of~$\xi^2$ with increasing grid size of $(N+N)\times N$ for the~$\T_H = 0.5$ ($\alpha \approx 0.407$) droplet.  Note the log-linear scale; the convergence is close to exponential.  \subref{subfig:convergencepower}: Convergence of~$\xi^2$ with increasing grid size of $(N+N)\times N$ for the Boulware~$\T_H = 0$ ($\alpha =1$) droplet.  The scale is now log-log; the convergence is power law with fitted power $N^{-2.5}$, which within our uncertainty agrees with the predicted behavior~$N^{-(\sqrt{11}-1)} \approx N^{-2.3}$.}
\label{fig:convergence}
\end{figure}


\section{Constructing Black Funnels}
\label{sec:constructionbf}

In the previous sections, we have constructed zero-temperature droplet solutions.  However, in order to determine whether the bulk dual to the Boulware vacuum is a droplet or a funnel, we must see if zero-temperature funnels exist. If they do, we must find which solution is thermodynamically preferred.  Thus in this section we construct black funnels with~$\T \equiv \T_\infty = \T_{\mathrm H}$ and try to push these temperatures to zero to obtain the desired Boulware vacuum.

\begin{figure}[t]
\centering
\subfigure[]{
\includegraphics[width=0.4\textwidth,page=7]{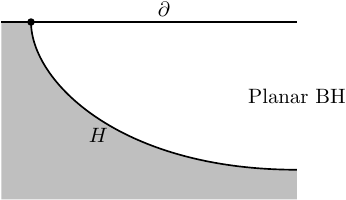}
\label{subfig:funneltriangle}
}
\hspace{1cm}
\subfigure[]{
\includegraphics[width=0.4\textwidth,page=8]{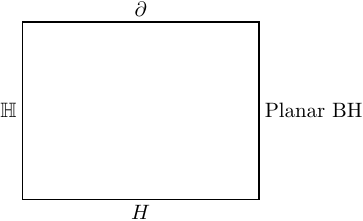}
\label{subfig:funnelsquare}
}
\caption{\subref{subfig:funneltriangle}: A sketch of a finite-temperature funnel.  The domain has three boundaries: the conformal boundary~$\partial$, the funnel horizon~$H$, and a planar black hole.  As for the droplets, if~$\T \neq 1$, the coordinate ``point'' where the bulk horizon meets the boundary is ill-behaved.  \subref{subfig:funnelsquare}: The ``point'' where the horizon meets the boundary can be blown up into a hyperbolic black hole~$\Hbb$.  This transformation regulates the singular point, and as an added benefit yields a four-sided computational domain.}
\label{fig:funneldomain}
\end{figure}

Let us therefore briefly outline the construction of such funnels.  As shown in Figure~\ref{subfig:funneltriangle}, a natural sketch of a funnel has three boundaries: the conformal boundary, the horizon, and a planar black hole.  However, just as for the droplets, in the detuned case~$\T \neq 1$ the coordinate ``point'' where the horizon meets the boundary is singular, and thus must be regulated by expanding it into an asymptotically hyperbolic black hole.  The resulting four-sided integration domain is shown in Figure~\ref{subfig:funnelsquare} (in fact, since this procedure yields a rectangular domain, it was used even in the Hartle-Hawking case~$\T = 1$ to simplify the numerics~\cite{SanWay12}).

We thus consider the following ansatz:
\begin{multline}
\label{eq:funnelansatz}
\mathrm{d}s^2 = \frac{\ell^2}{x y H^2}\left[-x (1-y) M Q_1\mathrm{d}t^2+\frac{x Q_2 H^2 y_+^2}{4y (1-y) M} \, \mathrm{d}y^2 \right. \\ \left. + \frac{Y_0^2 Q_4}{x(1-x)^4}\left(\mathrm{d}x+x(1-x)^2 Q_3\mathrm{d}y\right)^2+\frac{Y_0^2 Q_5}{(1-x)^2}\mathrm{d}\Omega_2^2\right],
\end{multline}
where
\begin{subequations}
\bea
G &= y_+^2-y(1-y_+^2), \\
H &= 2y_+^2 - 1 + x, \\
M &= G(1-x)+y_+^2 x(1+y), \\
Y_0 &= y_+ \left(y_+^2-\frac{1}{2} \right),
\eea
\end{subequations}
and the~$Q_i$ are unknown functions of~$x$ and~$y$ (we also note that these~$(x,y)$ are obviously not the same as those used in the construction of the black droplets above).  Here~$y_+$ is a parameter controlling~$\T$ via
\begin{equation}
\T=\frac{2y_+^2-1}{y_+};
\end{equation}
thus the ansatz above reduces to the Hartle-Hawking ansatz used in~\cite{SanWay12} when~$y_+ =1$.

The boundary conditions on~\eqref{eq:funnelansatz} are detailed in Appendix~\ref{subapp:funnel}.  In short, they are:
\begin{itemize}
	\item At the conformal boundary ($y = 0$), the hyperbolic black hole ($x = 0$), and the planar black hole ($x = 1$), we impose the Dirichlet conditions~$Q_1=Q_2=Q_4=Q_5=1$ and~$Q_3=0$.
	\item At the bulk horizon ($y = 1$), we impose regularity, requiring~$\partial_y Q_i = 0$ and~$Q_1 = Q_2$.
\end{itemize}
For a reference metric to use with the de Turck method, we take~\eqref{eq:funnelansatz} with~$Q_1=Q_2=Q_4=Q_5=1$ and $Q_3=0$.

Thus to detune the black funnel, we look for solutions to the Einstein-de Turck equations with the ansatz~\eqref{eq:funnelansatz}, tuning~$\T$ from unity towards zero.

\subsection{Nonexistence of Boulware Funnels}

\begin{figure}[t]
\centering
\subfigure[]{
\includegraphics[width=0.46\textwidth,page=15]{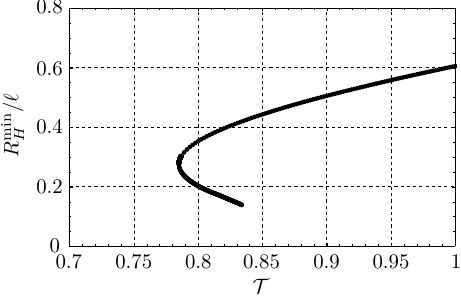}
\label{subfig:rmin}
}
\hspace{0.4cm}
\subfigure[]{
\includegraphics[width=0.46\textwidth,page=16]{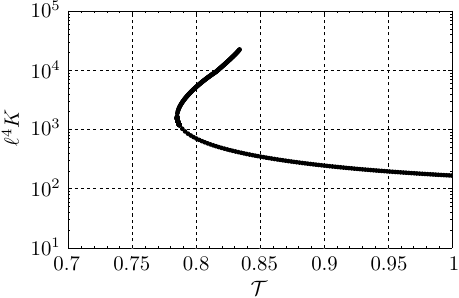}
\label{subfig:kret}
}
\caption{\subref{subfig:rmin}: The minimum areal radius~$R_H^\mathrm{min}$ of the funnels as a function of temperature.  Note that a minimum temperature of~$\T_\mathrm{min} \approx 0.785$ is reached where the thick (upper) and thin (lower) funnel branches meet.  We expect the minimum radius of the thin funnel branch to continue to decrease to zero, indicating a ``pinch-off'' of the funnel horizon.  \subref{subfig:kret}: The Kretschmann scalar~$K \equiv R_{abcd}R^{abcd}$ of the detuned funnels evaluated where the bulk horizon attains its minimal areal radius.  Note the logarithmic scale;~$K$ appears to diverge as the funnels pinch off.}
\label{fig:nofunnel}
\end{figure}

We now summarise our results.  We study the structure of our solutions by keeping track of the minimal areal radius~$R^\mathrm{min}_H$ of the funnel horizon.  Starting with the Hartle-Hawking ($\T = 1$) funnel, we may initially decrease~$\T$, and find that~$R^\mathrm{min}_H$ also decreases.  However, at~$\T \approx 0.785$ a minimum-temperature solution is reached.  It is possible to continue to decrease~$R^\mathrm{min}_H$ further, but only by \textit{increasing} ~$\T$, as shown in Figure~\ref{subfig:rmin}.  Therefore, it is possible for two different black funnels to exist at the same temperature. We distinguish these funnels by their shape by calling them ``thick funnels'' or ``thin funnels'' (the top and bottom branches of Figure~\ref{subfig:rmin}, respectively).  Crucially, extrapolation of the behavior shown in Figure~\ref{subfig:rmin} appears to imply that the thin funnel branch should reach~$R^\mathrm{min}_H = 0$, indicating a pinch-off transition towards a black droplet phase.

Indeed, as an additional check, in Figure~\ref{subfig:kret} we plot the Kretschmann scalar~$K$ evaluated on the funnel horizon~$\mathcal{H}$ where the minimum radius~$R^\mathrm{min}_H$ is attained; note that~$K$ appears to diverge as the thin funnels become narrower, consistent with the expectation that the funnel solutions become singular when~$R^\mathrm{min}_H = 0$.  Although our numerical methods cannot reach~$R^\mathrm{min}_H = 0$, our results provide strong evidence that no funnel phase with~$\T_H = \T_\infty = 0$ exists, and therefore that the Boulware vacuum is dual to a droplet phase.


\section{Stress Energy Tensors}
\label{sec:stresstensor}

\begin{figure}
\centering
\includegraphics[width = 0.6\textwidth,page=11]{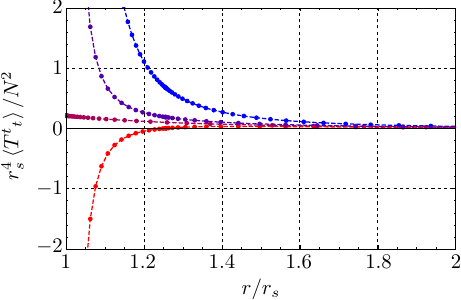}
\caption{The $\langle T^t{}_t\rangle$ component of the stress energy tensor dual to the hairless droplets as a function of $r/r_s$.  From bottom to top, the curves correspond to $\T_{\mathrm H} = 1.05, 1, 0.95$, and~$0$. $\mathcal T_{\mathrm H}=1$ is the Unruh state and $\mathcal T_{\mathrm H}=0$ is the Boulware state.  Note that the stress tensors of all detuned states diverge at the horizon.}
\label{fig:dropletstress}
\end{figure}

Let us now present the boundary stress tensors of the droplets and funnels described above, computed using standard holographic renormalization~\cite{BalKra99,deHSol00}.  In Figure~\ref{fig:dropletstress}, we show the~$\langle T^t{}_t\rangle$ component of the stress tensor for the hairless ($\varphi = 0$) black droplets, with varying~$\T_H \in [0,1.05]$.  For all of these solutions,~$\langle T^t{}_t\rangle$ has an asymptotic falloff of~$r^{-5}$.  Moreover, recall that~$\langle T^t{}_t\rangle$ has the opposite sign from the (static) local energy density~$\rho = \langle T_{tt}\rangle$.  We thus immediately see from the figure that the $\T_H = 0$ droplet has the lowest pointwise energy density in this family.  In fact, it also has lower pointwise energy density than the $\mathcal T_\infty\neq0$ droplets in \cite{SanWay14}, as well as the black funnels (to be presented shortly).  As expected, in the universal sector of AdS/CFT (i.e.~with no bulk matter fields), the~$\T_H = \T_\infty = 0$ droplet is therefore likely dual to the Boulware state: that is, it is dual to the CFT state on Schwarzschild of lowest total energy.  Note also that this energy density is everywhere negative, and in fact it can be shown that all of the classical energy conditions are violated everywhere.

\begin{figure}
\centering
\includegraphics[width = 0.6\textwidth,page=12]{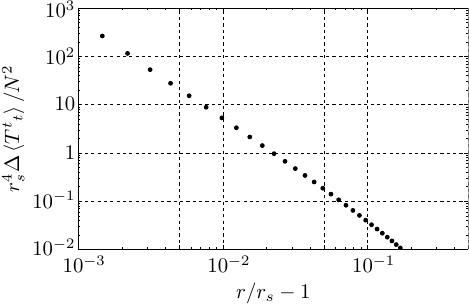}
\caption{$\Delta \langle T^t{}_t\rangle$ for the $\mathcal T_{\mathrm H}=0.025$ hairy droplet as a function of $r/r_s$.  Here we show $\Delta \langle T^t{}_t \rangle$ down to about the precision with which we are able to numerically extract it; at larger~$r$, it becomes zero within the precision of our extraction.  Note that it is everywhere non-negative, indicating that this solution has a lower local energy density (and thus total energy) than the hairless Boulware droplet.}
\label{fig:hairydropletstress}
\end{figure}

Next, in Figure~\ref{fig:hairydropletstress} we compare the stress tensor of a hairy solution to that of the vacuum Boulware droplet by plotting
\be
\label{eq:Tdiffhairy}
\Delta \langle T^t{}_t\rangle \equiv \langle T^t{}_t\rangle_\mathrm{hairy} - \langle T^t{}_t\rangle_{\mathrm{hairless} \, \mathrm{Boulware}}.
\ee
The particular solution shown in Figure~\ref{fig:hairydropletstress} has $\T_H = 0.025$, which is the lowest temperature we have reached, though we note that extrapolation of our data implies that a hairy solution at~$\T_H = 0$ should indeed exist.  However, even without taking~$\T_H$ to zero, we observe that~$\Delta \langle T^t{}_t\rangle$ is positive, and thus even this finite-temperature hairy droplet has lower energy than the zero-temperature hairless droplet (taking the hairy droplet all the way to zero temperature should just lower its energy further).  This illustrates that the Boulware state is theory-dependent.  Specifically, in the universal sector of AdS/CFT, the Boulware vacuum is just the hairless droplet described above.  But by introducing a bulk scalar field (dual to a scalar CFT operator), the Boulware state becomes the zero-temperature limit of the hairy droplet shown in Figure~\ref{fig:hairydropletstress}, which has a lower energy than its hairless counterpart.

\begin{figure}
\centering
\includegraphics[width=0.6\textwidth,page=14]{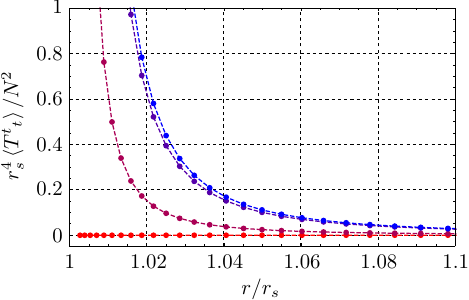}
\caption{The $\langle {T^t}_t\rangle$ component of the stress energy tensor of the detuned funnels as a function of $r/r_s$.  From bottom to to top, the curves show~$\T \approx 1$,~0.956,~0.812, and~0.785, with corresponding minimal radii~$R^\mathrm{min}_H \approx 0.605$,~0.56,~0.178, and~0.277 (note therefore that the bottom two curves are on the thick funnel branch, while the second curve from the top lies on the thin funnel branch).  Note that the~$\T = 1$ funnel corresponds to the Hartle-Hawking vacuum, and one can indeed check that~$\langle {T^t}_t\rangle$ is negative on the lowermost curve (though it cannot obviously be inferred from the plot due to scaling).}
\label{fig:funnelstress}
\end{figure}

Finally, in Figure~\ref{fig:funnelstress} we show the stress tensor of the funnels; as for the detuned droplet, all the detuned funnels exhibit stress tensors that are singular at the horizon.  Note that the energy density for the tuned funnel ($\T = 1$), corresponding to the Hartle-Hawking vacuum, is everywhere positive, as would be expected in a state of global thermal equilibrium.  However, like the droplets, all the detuned funnels exhibit a negative energy density near the horizon.

\section{Discussion}
\label{sec:discussion}

In this paper, we have constructed the gravitational dual to the Boulware state of a holographic CFT on the Schwarzschild spacetime.  This dual is an extremal black droplet, a zero-temperature black hole anchored to the AdS boundary.  We emphasise that droplets with~$\T_H = \T_\infty = 0$ and~$T_\mathrm{BH} \neq 0$ have never been constructed before, and thus the extremal droplet presented here provides the first exploration of Boulware states of strongly interacting field theories on black hole spacetimes.  

In order to confirm that the gravitational dual to the Boulware state is a droplet, we have also constructed detuned black funnels, that is, funnels with~$\T \equiv \T_H = \T_\infty \neq T_\mathrm{BH}$.  Our results suggest that as~$\T$ is decreased, the funnels ``pinch off'' at nonzero~$\T$, which we interpret as evidence that a funnel with~$\T = 0$ does not exist. This implies that the bulk dual to the Boulware state is indeed the droplet.

We should note, however, that the Boulware state is of course theory-dependent.  For instance, we have shown that if we introduce a scalar field in the bulk, at sufficiently low temperature the scalar field condenses around the black droplet.  The resulting hairy black droplet has a lower energy than the hairless Boulware droplet, so for a theory containing this bulk scalar field, the bulk dual to the Boulware state is the zero-temperature \textit{hairy} black droplet.  In particular, this implies that changing the theory allows us to obtain states with lower total energy than that of the hairless black droplet.

In fact, the presence of negative energies leads to an interesting observation.  Note that much as in free field theory calculations on Schwarzschild spacetime, the (hairless) Boulware state we have found here exhibits a negative energy density as measured by a static observer.  It is straightforward to check that it violates the null energy condition
\be
\left\langle T_{ab} k^a k^b \right\rangle \geq 0 \mbox{ for all null } k^a,
\ee
and thus it must violate all the standard classical local positive-energy conditions as well.  This violation is unsurprising, as it is a general property of QFTs~\cite{EpsGla65}.  What is more interesting is that when compared to other static states (e.g.~the Unruh and Hartle-Hawking states), the energy density of the Boulware state is more negative.  This is reminiscent of so-called quantum energy inequalities (QEIs)~\cite{PfeFor97,Fla97,FewTeo98,Few99,Fla02,FewHol04,FewOst07,FewSmi07,Few12}, which in certain contexts constrain violations of classical energy conditions relative to some reference background state\footnote{Though note that there also exist so-called ``absolute QEIs'' which place bounds on the renormalized expectation value of the energy density~\cite{FewSmi07}.}.  These ``difference" QEIs often take the form
\be
\label{eq:Tdiff}
\left\langle T_{ab} \right\rangle_\mathrm{diff} \equiv \left\langle T_{ab} \right\rangle - \left\langle T_{ab} \right\rangle_\mathrm{background}\geq -B\;,
\ee
where $\left\langle T_{ab} \right\rangle_\mathrm{background}$ is the stress tensor for some background state and $B$ is some non-negative quantity that is a function only of this background state.  For constructing QEIs for free field theories on the Schwarzschild spacetime~\cite{PfeFor97,FewTeo98}, this background state is typically taken to be the Boulware state.  It would therefore be interesting to use the Boulware state constructed here to examine whether or not the background-subtracted energy density obeys similar bounds.  We emphasise that such bounds are only sensible when comparing states of the \textit{same} theory.  For instance, one might be tempted use the difference~\eqref{eq:Tdiffhairy} between the hairy and hairless Boulware droplets to test or construct these QEIs.  But this comparison is only allowed if one thinks of the hairless droplet as an excited state in a theory containing the bulk scalar~$\varphi$.  Then since the CFT dual of the hairless droplet has a positive energy density relative to that of the hairy droplet, one doesn't obtain any interesting statements about energy negativity relative to the vacuum.

It would be interesting to study gravitational perturbations of the Boulware state to see if~$\left\langle T_{ab} \right\rangle_\mathrm{diff}$ can become negative, thus providing some guidance towards developing nontrivial QEIs.  While such an analysis is outside the scope of the present work, to our knowledge any such bounds would be the first for a strongly-interacting field theory in greater than two dimensions (two-dimensional CFTs were shown to obey a set of QEIs in~\cite{FewHol04}), though we note that a partial result in this direction will appear shortly~\cite{FisWis16}.  We leave this as an avenue of future investigation.

We finish by summarising the state of affairs for the holographic Unruh, Hartle-Hawking, and Boulware states on a Schwarzschild background.  Since droplets and funnels have different transport properties, it is desirable to know whether each of these states is dominated by a droplet or funnel solution. With our results, the Boulware state is likely represented by a droplet. Black funnels representing the Hartle-Hawking state were found in \cite{SanWay12}, and later results \cite{SanWay14} suggest that the corresponding droplet solutions do not exist. Droplets for the Unruh state were found in \cite{FigLuc11} (and reproduced in \cite{SanWay14}), but it remains unclear whether funnel solutions exist. One means of attempting to find such a solution would be to begin with the Hartle-Hawking funnel, and lower $\mathcal T_\infty$ until it vanishes. Because these black funnels contain a connected horizon with different asymptotic temperatures, their horizon would be non-Killing like those in \cite{FisMar12b,FigWis12,EmpMar13}. 

\acknowledgments

It is a pleasure to thank Chris Fewster and Donald Marolf for useful discussions and correspondence.  SF is supported by the ERC Advanced grant No.~290456 and by STFC grant ST/L00044X/1.  SF also thanks the University of California Santa Barbara for hospitality during stages of this work. BW was  supported by European Research Council grant no. ERC-2011-StG 279363-HiDGR and NSERC.

\appendix

\section{Detuning Boundary and Bulk Temperatures}
\label{app:detuning}

Here let us briefly review the technique introduced in~\cite{FisMar12} for detuning~$T_\mathrm{BH}$ and~$T_H$.  First, consider a conformal transformation of the Schwarzschild metric~\eqref{eq:schwarzschild}, $\dd s^2_{\mathrm{Schw}}\rightarrow \dd s^2_{\mathrm{Schw}}/f(r)$, followed by a coordinate transformation from~$r$ to~$z=2r_s \sqrt{f}/r$.  Near the horizon~$r=r_s$ ($z=0$), the metric becomes
\be
\frac{\dd s^2_{\mathrm{Schw}}}{f(r)} = -\dd t^2+4r_s^2\left(\frac{\dd z^2}{z^2}+\frac{\dd\Omega_2^2}{z^2}\right) + \mathcal{O}(z^0).
\ee
Neglecting the subleading terms in~$z$, the above geometry is~$\mathbb R_t \times \Hbb^3$, where $\Hbb^d$ is $d$-dimensional Euclidean hyperbolic space.  This hyperboloid $\Hbb^3$ has a length scale $\ell_{\mathrm{hyp}}=2r_s$.  Thus, in this conformal frame (often called the \emph{ultrastatic} frame), the Schwarzschild horizon has been replaced by an asymptotically hyperbolic region.  From this perspective, it is clear that one can place a thermal bath at any temperature~$T_H$ at such an asymptotic region (and not just~$T_\mathrm{BH}$).

Such a heat bath will be dual to a bulk black hole with hyperbolic symmetry: namely, the hyperbolic Schwarzschild-AdS black hole~\cite{Emp99}:
\be
\label{eq:hyperbolic}
\dd s^2_{\Hbb} = -g(r)\frac{\ell^2}{\ell_{\mathrm{hyp}}^2}\dd t^2+\frac{\dd r^2}{g(r)}+r^2 \dd\Sigma_3^2, \mbox{ with } g(r) = \frac{r^2}{\ell^2}-1-\frac{r_0^2}{r^2}\left(\frac{r_0^2}{\ell^2}-1\right).
\ee
Here~$\dd\Sigma_3=\dd\eta^2+\sinh^2\eta\,\dd\Omega_2^2$ is the metric on the unit Euclidean hyperboloid and~$r_0$ is a free parameter that sets the bulk horizon temperature as
\be
T_H = \left(\frac{2r_0}{\ell} - \frac{\ell}{r_0}\right) T_\mathrm{BH}.
\ee
Thus by requiring that the bulk metric approach~\eqref{eq:hyperbolic} in an asymptotic region, we may choose the bulk horizon temperature~$T_H$ arbitrarily, and in particular may take it to be different than the boundary black hole temperature~$T_\mathrm{BH}$.

Note that in practice, we find it more convenient to work with the coordinates $x=1-\sqrt{1-r_0^2/r^2}$, $y=\sqrt{1-\mathrm{csch}\,\eta}$, in terms of which the metric becomes
\begin{multline}
\label{eq:hyperbolicbh}
\dd s^2_{\Hbb} = \frac{\ell^2}{x(2-x)(1+\alpha)}\left[-\frac{(1-x)^2(1-\alpha x(2-x))}{4r_s^2}\, \dd t^2 \right. \\ \left. +\frac{(1+\alpha)}{x(2-x)(1-\alpha x(2-x))}\, \dd x^2+\frac{4y^2\,\dd y^2}{(1-y^2)^2(1+(1-y^2)^2)}+\frac{\dd\Omega_2^2}{(1-y^2)^2}\right],
\end{multline}
where we have defined~$\alpha \equiv \ell^2/r_0^2-1$ and we have substituted $\ell_{\mathrm{hyp}}=2r_s$.  In these coordinates, the asymptotically hyperbolic region is at~$y = 1$ and the temperature can be expressed as~\eqref{eq:hypertemp} in the main text.

\section{Boundary Conditions}
\label{app:bndry}

In this Appendix, we provide additional details on the boundary conditions used to obtain the droplets and funnels presented in the main text.

\subsection{Droplet}
\label{subapp:droplet}

Let us show that the Dirichlet boundary conditions on the~$Q_i$ listed in Section~\ref{subsec:ansatz} give the correct boundary conditions on the metric.  First, near the conformal boundary~($x = 0$), the ansatz~\eqref{eq:ansatz} takes the form
\be
\dd s^2|_{x \rightarrow 0}=\frac{\ell^2}{x}\left[\frac{\dd x^2}{4x}+\frac{2y^4(2-y^2)^2}{r_s^2(1+\alpha)(1-y^2)^2(1+y\sqrt{2-y^2})^2} \, \dd s_\mathrm{Schw}^2\right],
\ee
where~$\dd s_\mathrm{Schw}^2$ is the Schwarzschild line element in the form
\be
\dd s_\mathrm{Schw}^2 =-\frac{(1-y\sqrt{2-y^2})^2}{(1-y^2)^2} \, \dd t^2 + \frac{r_0^2(1+y\sqrt{2-y^2})^2}{4y^2(2-y^2)}\left(\frac{4 \, \dd y^2}{y^2(2-y^2)^2}+\dd\Omega_2^2\right).
\ee
This can be put into the more familiar form~\eqref{eq:schwarzschild} by the coordinate transformation
\be
y = \sqrt{1+\frac{2r\sqrt{1-r_s/r}}{r_s - 2r}}.
\ee

Next, near the hyperbolic black hole~($y = 1$), the ansatz~\eqref{eq:ansatz} becomes
\begin{multline}
\dd s^2|_{y\rightarrow 1} =\frac{\ell^2}{x(2-x)(1+\alpha)}\left[-\frac{(1-x)^2(1-\alpha x(2-x))}{4r_s^2} \, \dd t^2 \right. \\ \left. +\frac{(1+\alpha)\,\dd x^2}{x(2-x)(1-\alpha x(2-x))}+\frac{4\,\dd y^2}{(1-y^2)^2}+\frac{\dd\Omega_2^2}{(1-y^2)^2}\right],
\end{multline}
which is precisely the $y \rightarrow 1$ asymptotic region of the hyperbolic black hole \eqref{eq:hyperbolicbh}.

The Poincar\'e horizon lies at the coordinate point $(x,y) = (0,0)$, corresponding to the boundary~$\rho = 0$ in the~$(\rho, \chi)$ coordinates defined by~\eqref{eq:coordtrans}.  Re-expressing~\eqref{eq:ansatz} in these coordinates, we find that near~$\rho = 0$, the ansatz becomes
\be
\dd s^2|_{\rho\rightarrow 0}=\frac{\ell^2}{1-\chi^2}\left(-\frac{8\rho^2\,\dd t^2}{r_s^2}+\frac{\dd\rho^2}{\rho^2}+\frac{\dd\chi^2}{1-\chi^2}+\chi^2\dd\Omega_2\right),
\ee
which agrees with the Poincar\'e metric~\eqref{eq:poincare} up to a trivial rescaling of time.

Finally, regularity of the bulk horizon~($x = 1$) and the axis of symmetry~($y = 0$) follows from the Neumann and Dirichlet boundary conditions imposed there; see e.g.~\cite{DiaSan15}.

\subsection{Funnel}
\label{subapp:funnel}

Let us show that the Dirichlet boundary conditions on the~$Q_i$ listed in Section~\ref{sec:constructionbf} give the correct boundary conditions on the metric.  First, near the conformal boundary~($y = 0$), the ansatz~\eqref{eq:funnelansatz} approaches
\be
\mathrm{d}s^2|_{y \to 0} = \frac{\ell^2}{y}\left[\frac{H}{4y} \, \dd y^2 + \frac{Y_0^2}{r_s^2 x H}  \, \dd s_\mathrm{Schw}^2 \right],
\ee
where the Schwarzschild line element appears in the form
\be
\mathrm{d}s_\mathrm{Schw}^2 = -x \frac{r_s^2 y_+^2}{Y_0^2} \, \mathrm{d}t^2+\frac{r_s^2}{x(1-x)^4} \, \mathrm{d}x^2 + \frac{r_s^2}{(1-x)^2}\mathrm{d}\Omega_2^2.
\ee
This can be reduced to the familiar form~\eqref{eq:schwarzschild} via the coordinate redefinitions~$r = r_s/(1-x)$ and~$t \to (r_s Y_0/y_+) t$.

Next, near the hyperbolic black hole~($x = 0$), the ansatz~\eqref{eq:funnelansatz} approaches
\be
\dd s^2|_{x \to 0} = \frac{\ell^2 y_+^2}{4y}\left[-\frac{(1-y)G(y)}{Y_0^2} \, \mathrm{d}t^2+\frac{\mathrm{d}y^2}{y (1-y) G(y)}+\frac{\mathrm{d}x^2}{x^2}+\frac{1}{x}\mathrm{d} \, \Omega_2^2 \right],
\ee
which coincides with the large-$\eta$ limit of~\eqref{eq:hyperbolic} if we identify $t = (\ell/\ell_{\mathrm{hyp}}) G(1) t_{\mathrm{hyp}}$, $r = r_0/\sqrt{y}$, and $r_0 = y_+ \ell$.

At the planar black hole ($x = 1$), the ansatz approaches
\be
\mathrm{d}s^2|_{x \to 1} = \frac{\ell^2}{y}\left[-\frac{(1-y^2)}{4 y_+^2}\mathrm{d}t^2+\frac{\mathrm{d}y^2}{4y (1-y^2)}+\frac{Y_0^2\mathrm{d}x^2}{(1-x)^4}+\frac{Y_0^2}{(1-x)^2}\mathrm{d}\Omega_2^2 \right],
\ee
which takes the usual form of a planar Schwarzschild black brane if we transform to new coordinates~$y=z^2$,~$r = Y_0/(1-x)$ and $t \to 2y_+ t$. 

Finally, as for the droplet, regularity of the bulk horizon~($y = 1$) in ingoing Eddington-Finkelstein coordinates follows from the Neumann and Dirichlet boundary conditions imposed there.

\section{Convergence of the Boulware Droplet}
\label{app:conv}
As we have shown in Figure~\ref{fig:convergence}, our non-extremal droplet solutions converge exponentially, while the extremal Boulware droplet only converges at a power law.  In this Appendix, we demonstrate that this behaviour is in accordance with the expectations of pseudospectral methods. 

In general, smooth functions are expected to exhibit exponential convergence. Indeed, non-extremal droplets (with~$\alpha \neq 1$) have finite-temperature Killing horizons, which can be shown to give rise to regular singular points (i.e.~to analytic metric components).  As a result the convergence of these solutions should have exponential convergence, as verified in Figure~\ref{subfig:convergenceexp}.  

However, non-smooth behaviour can arise in the Boulware droplet ($\alpha = 1$) for which the bulk horizon is extremal.   In general, if a function is $C^p$ but not $C^{p+1}$, pseudo-spectral methods on a Chebyshev grid with $N$ points will converge to the continuum limit as $N^{-p}$~\cite{opac-b1127559}.  In order to obtain the expected convergence on the Boulware droplet, we therefore need to quantify the differentiability of its metric components (in the coordinates of Section~\ref{sec:construction}).

For the Boulware droplet, we have verified that it is consistent to impose Dirichlet conditions directly on the extremal horizon~$H$, which implies that it will exhibit the same analytic behaviour as the hyperbolic black hole~$\mathbb{H}$.  We therefore expect the convergence of the method to be dictated by the behaviour of normalisable and static perturbations around the extremal hyperbolic black hole.

Let us therefore consider stationary and spherically symmetric perturbations of the extremal hyperbolic black hole: we wish to perturb
\be
\label{eq:backap}
\mathrm{d}s^2=-f(r)\mathrm{d}t^2+\frac{\mathrm{d}r^2}{f(r)}+r^2(\mathrm{d}\tau^2 +\sinh^2\tau \mathrm{d}\Omega_2^2),
\ee
where
\be
f(r) = \frac{r^2}{\ell^2}-1-\frac{r_0^2}{r^2}\left(\frac{r^2_0}{\ell^2}-1\right),
\ee
which at zero temperature is
\be
f(r) = \frac{\ell^2}{r^2} \left(\frac{r^2}{\ell^2} - \frac{1}{2}\right)^2.
\ee

The most general perturbations respecting stationarity and $SO(3)$ symmetry take the form
\be
\delta \mathrm{d}s^2\equiv h_{ab}\mathrm{d}x^a\mathrm{d}x^b=-f(r)q_1\mathrm{d}t^2+\frac{q_2\mathrm{d}r^2}{f(r)}+r^2(q_3\mathrm{d}\tau^2+q_4\sinh^2\tau \mathrm{d}\Omega_2^2)+q_5 \mathrm{d}\tau \mathrm{d}r,
\ee
where $q_1,\,q_2,\,q_3,\,q_4,\,\text{and}\,q_5$ depend on $r$ and $\tau$ only.  Since we have two degrees of freedom associated with independent infinitesimal coordinate transformations of $\tau$ and $r$, we can choose the so-called spherical gauge in which $q_5=0$ and $q_4=q_3$. This brings the metric perturbations to the following form:
\be
\delta \mathrm{d}s^2\equiv h_{ab}\mathrm{d}x^a\mathrm{d}x^b=-f(r)q_1\mathrm{d}t^2+\frac{q_2\mathrm{d}r^2}{f(r)}+r^2q_3(\mathrm{d}\tau^2+\sinh^2\tau \mathrm{d}\Omega_2^2).
\ee
To proceed further, we take advantage of the symmetry of the background solution~\eqref{eq:backap}.  The isometry group of~\eqref{eq:backap} is $\mathbb{R}_t\times SO(1,3)$, and thus we can decompose our static perturbations according to how they transform under~$SO(1,3)$.  Specifically, gravitational perturbations about~\eqref{eq:backap} will come in three classes, which transforms as tensors, vectors, and scalars in $SO(1,3)$.  It can be shown that the requirement that~$SO(3)$ be preserved within the~$SO(1,3)$ forbids the tensor and vector perturbations, and thus we are left with scalars.

These scalar gravitational perturbations are then constructed from scalar harmonics on $\mathbb{H}_3$ as
\be
\label{eq:scalardecomp}
q_i(r,\tau) = \hat{q}_i(r) H(\tau)\,,\text{for}\,i\in\{1,2,3\},
\ee
where due to the assumption of~$SO(3)$ symmetry we only consider those harmonics~$H$ that depend only on~$\tau$.  Such harmonics therefore obey
\be
\Box_{\mathbb{H}_3} H+\lambda H=0 \quad \Rightarrow \quad \frac{1}{\sinh^2\tau}\frac{\mathrm{d}}{\mathrm{d}\tau}\left(\sinh^2\tau \frac{\mathrm{d}H}{\mathrm{d}\tau}\right) +\lambda H = 0,
\ee
whose general solution is
\be
H (\tau) = -\frac{1}{2i\,\sinh\tau}\left(e^{-\sqrt{1-\lambda }\,\tau } C_1+e^{\sqrt{1-\lambda }\,\tau } C_2\right).
\ee
Normalizability at~$\tau = 0$ requires~$C_2 = -C_1$, while normalizability at large~$\tau$ requires~$\lambda = 1+\Lambda^2$ for $\Lambda\in\mathbb{R}$; thus
\be
H (\tau)=C_1\frac{\sin (\Lambda \tau)}{\sinh \tau}.
\ee

Inserting the decomposition~\eqref{eq:scalardecomp} into the Einstein equation allows us to express $\hat{q}_1$ and $\hat{q}_2$ as a function of $\hat{q}_3$ and its first derivative:
\begin{subequations}
\bea
\hat{q}_1 &= -\hat{q}_2-\hat{q}_3, \\
\hat{q}_2 &= \frac{\left[\ell^2(\lambda +3) r^2 + 6 r_0^2(\ell^2 -r_0^2)\right] \hat{q}_3 - 3 r \left(r^4+r_0^4-\ell^2 r_0^2\right) \hat{q}_3'}{\lambda \ell^2 r^2 + 3(r^2 - r_0^2)(\ell^2 - r^2 - r_0^2)},
\eea
\end{subequations}
where~$\hat{q}_3' \equiv \mathrm{d}\hat{q}_3/\mathrm{d}r$.  We also find that~$\hat{q}_3$ obeys the following second order differential equation:
\begin{multline}
\label{eq:q3hat}
\hat{q}_3'' + \frac{1}{r}\left[2 r^2 \left(\frac{1}{r^2+r_0^2-\ell^2}+\frac{1}{r^2-r_0^2}\right)+\frac{6 \left(r^4+r_0^4-\ell^2 r_0^2\right)}{\ell^2 \left[(\lambda +3) r^2-3 r_0^2\right]-3 r^4+3 r_0^4}+1\right]\hat{q}_3' \\
+\frac{\lambda  \ell^2 }{\left(r^2-r_0^2\right) \left(r^2+r_0^2-\ell^2\right) }\frac{\ell^2 \left[(\lambda +3) r^2+3 r_0^2\right]-r^4-3 r_0^4}{3 \left(r^4-r_0^4\right)-\ell^2 \left[(\lambda +3) r^2-3 r_0^2\right]}\hat{q}_3 = 0.
\end{multline}
Next, recall that we are only interested in the extremal limit~$r_0/\ell = 1/\sqrt{2}$.  In order to understand the behaviour close to the extremal horizon, we use Frobenius's method and investigate solutions of the form
\be
q_3(r) = \left(\frac{r}{r_0}-1\right)^s \sum_{i=0}^{+\infty}\left(\frac{r}{r_0}-1\right)^i a_i;
\ee
inserting this expansion into~\eqref{eq:q3hat}, we obtain the two allowed values of~$s$:
\be
s_{\pm} = \frac{1}{2} \left(\pm\sqrt{9+2 \lambda}-1\right).
\ee

Normalisability then requires us to discard the negative square root, and thus we have
\be
s = \frac{1}{2} \left(\sqrt{11+2 \Lambda^2}-1\right),
\ee
where we have also used the relation between $\lambda$ and $\Lambda$.  We thus find that with respect to the coordinate~$r$, all perturbations are $C^{\frac{1}{2}(\sqrt{11}-1)}$ or higher (since $\Lambda$ is real). To obtain the result quoted in the main text, we recall that near the horizon and exactly at extremality, the variable~$x$ that we used in our numerics relates to $r$
\be
r-r_0\sim (1-x)^2.
\ee
We thus find that the convergence to the continuum is limited to be $N^{-(\sqrt{11}-1)}\approx N^{-2.32}$.

\bibliographystyle{jhep}
\bibliography{biblio}

\end{document}